\documentclass[journal=jacsat,manuscript=letter,layout=onecolumn]{achemso}

\usepackage{achemso}
\setkeys{acs}{usetitle=true}
\usepackage{graphicx}
\usepackage[space]{grffile}
\usepackage{latexsym}
\usepackage{textcomp}
\usepackage{longtable}
\usepackage{multirow,booktabs}
\usepackage{amsfonts,amsmath,amssymb}
\usepackage{url}
\usepackage{hyperref}
\usepackage{color}

\hypersetup{colorlinks=false,pdfborder={0 0 0}}
\newif\iflatexml\latexmlfalse
\DeclareGraphicsExtensions{.pdf,.PDF,.png,.PNG,.jpg,.JPG,.jpeg,.JPEG}
\usepackage[version=3]{mhchem} 

\newcommand{\fig}[1]{Fig.~\ref{#1}}

\newcommand{\sige}[0]{{Si$_{1-x}$Ge$_x$~}}

\title[]{Dangling bonds as possible contributors to charge noise in silicon and silicon-germanium quantum dot qubits}
\author{Joel B. Varley}
\affiliation{Materials Science Division, Lawrence Livermore National Laboratory, Livermore, CA 94550, USA}
\email{varley2@llnl.gov}
\author{Keith G. Ray}
\affiliation{Materials Science Division, Lawrence Livermore National Laboratory, Livermore, CA 94550, USA}
\author{Vincenzo Lordi}
\affiliation{Materials Science Division, Lawrence Livermore National Laboratory, Livermore, CA 94550, USA}
\email{lordi2@llnl.gov}
\begin{document}

\newcommand{\td}[1]{{\bf TODO: #1 }}
%

\begin{abstract}



Spin qubits based on Si and \sige quantum dot architectures exhibit among the best coherence times of competing quantum computing technologies, yet they still suffer from charge noise that limit their qubit gate fidelities. 
Identifying the origins of these charge fluctuations is therefore a critical step toward improving Si quantum-dot-based qubits.
Here we use hybrid functional calculations to investigate possible atomistic sources of charge noise, focusing on charge trapping at Si and Ge dangling bonds (DBs). 
We evaluate the role of global and local environment in the defect levels associated with DBs in Si, Ge, and \sige alloys, and consider their trapping and excitation energies within the framework of configuration coordinate diagrams. 
We additionally consider the influence of strain and oxidation in charge-trapping energetics by analyzing Si and Ge$_{\rm Si}$ DBs in SiO$_2$ and strained Si layers in typical \sige quantum dot heterostructures.
Our results identify that Ge dangling bonds are more problematic charge-trapping centers both in typical \sige alloys and associated oxidation layers, and they may be exacerbated by compositional inhomogeneities. 
These results suggest the importance of alloy homogeneity and possible passivation schemes for DBs in Si-based quantum dot qubits and are of general relevance to mitigating possible trap levels in other Si, Ge, and Si$_{1-x}$Ge$_{x}$-based metal-oxide-semiconductor stacks and related devices.

\end{abstract}

\maketitle
%
\section{Introduction}

Of the many candidates for spin-based qubits, quantum dot architectures using Si-based materials are attractive owing to their long-coherence times, scalability, and 
well established fabrication technologies.\cite{Loss:1998gz,Kane:1998be,Zwanenburg:2013jr,Awschalom:2013ix,Veldhorst:2015dz} 
These devices typically involve laterally-patterned gates on Si/\sige or Si/SiO$_2$ (metal-oxide-semiconductor, or MOS) heterostructures that further can be isotopically purified to improve the already favorably-weak spin-orbit and hyperfine couplings that enable longer coherence times compared to other material choices.\cite{Zwanenburg:2013jr,Veldhorst:2015dz,Yoneda:2017cp}
Designs based on Si/\sige quantum wells have  demonstrated single-qubit gate fidelities exceeding 99.9\%, which is critical for scaling up fault-tolerant multi spin-qubit systems.\cite{Yoneda:2017cp}
Two qubit gate fidelities exceeding 94\% have been demonstrated in Si MOS devices.\cite{HuangNature2019}
Recent work has identified that 1/$f$ charge noise can become the dominant source of free-evolution dephasing of a single spin in Si-based quantum dots that exploit spin--electric field coupling and isotopic engineering, with qubit performance limited by electrical stability.\cite{Yoneda:2017cp} 
While the origins of charge noise can be attributed to a number of different mechanisms, charge traps near the heterostructure interfaces are widely believed to be the dominant sources.\cite{Fleetwood:1993hf,Paladino:2014dh,Bermeister:2014bv,Freeman:2016ca,Yoneda:2017cp,Connors:2019ej}
For example, recent experiments have provided a strong link between measured charge noise and defects in the vicinity of the semiconductor surface and/or oxide gate dielectric layer in Si/\sige quantum dot devices,\cite{Connors:2019ej} yet the exact defect identities remain unclear.

Here, we use  first-principles calculations based on screened hybrid density functional theory to explore the role of dangling-bond type defects in relevant layers of Si-based quantum dot qubit devices to investigate possible sources of charge noise in conventional Si/\sige quantum dot architectures.
Specifically, we investigate the energies of localized states associated with Si and Ge dangling bonds (DBs) in the strained Si layer, as well as how the DB energies may change with composition, which are of relevance to a number of \sige layers in typical devices (e.g. barrier layers in the vicinity of the quantum well) that may exhibit variations in the local composition.
Additionally, we consider Si and Ge DBs in SiO$_2$ as a proxy for these energies in a thermally oxidized layer or in the vicinity of the surface.
We compute defect formation energies ($E^f$), from which we can derive the stability of different charge states of each defect for the different materials and conditions (i.e., strain and composition) we consider, as well as the related electronic transition levels ($\varepsilon$) associated with ionization or charge-trapping processes that may occur in qubit devices under operation.\cite{Freysoldt:2014in}
We additionally evaluate energy barriers  associated with instantaneous charge-transfer between defect states, as may be expected for optical excitations or tunneling processes that can induce high- or low-frequency charge noise.

\begin{figure}
\includegraphics[width=1\columnwidth]{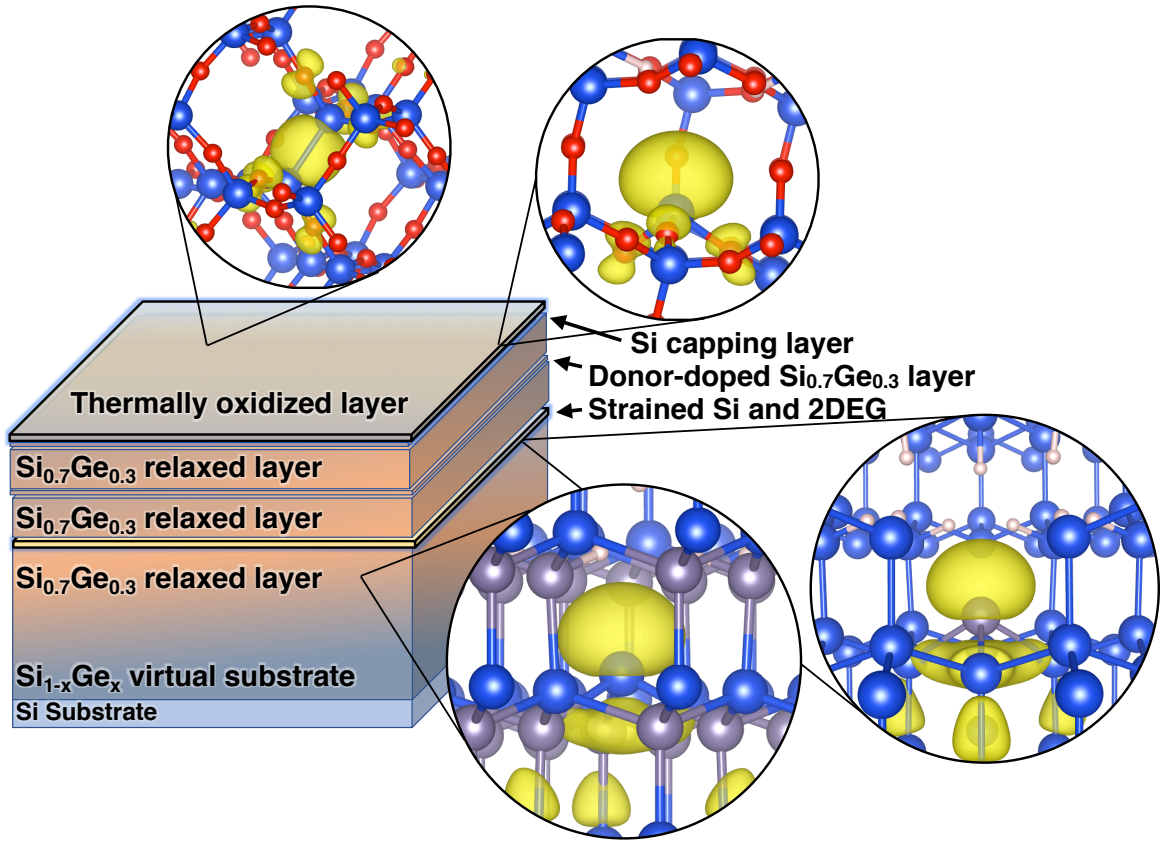}
\caption{Schematic diagram of a typical Si/\sige quantum dot material layer stack that adopts a strained Si quantum well grown between relaxed Si$_{0.7}$Ge$_{0.3}$ layers along the (001) direction and upper \sige layers that act as insulating barriers.\cite{Thalakulam:2010ef} 
Alternative architectures may include metal oxide semiconductor (MOS) designs with an oxidize layer  grown directly on the Si layer to act as insulator/barrier and replacing the upper \sige layers. Additional layers that may be present in certain device designs, such as Si capping layers and donor-doped layers to facilitate the formation of a two-dimensional electron gas (2DEG) are also depicted, but may not all be present in every device design. 
Additional patterned metal gates on top of the heterostructure stack and possible mesa-etched structures are omitted for clarity.\cite{Zwanenburg:2013jr}
Illustrations of various dangling-bond motifs that may form at hetero-interfaces, in the vicinity of extended defects (e.g. dislocations), or surfaces are included as insets in the various layers of the device. 
In the insets, Si atoms are blue, Ge atoms are grey, O atoms are red, and H atoms are pink, with charge-density iso-surfaces for the localized states associated with negatively-charged dangling bond states shown in yellow at 10\% of their maximum values. Electron tunneling and/or charge-transfer from these states may contribute to charge noise observed in these devices. 
} \label{FIG:DIAGRAM}
\end{figure}

\begin{figure}
\includegraphics[width=1\columnwidth]{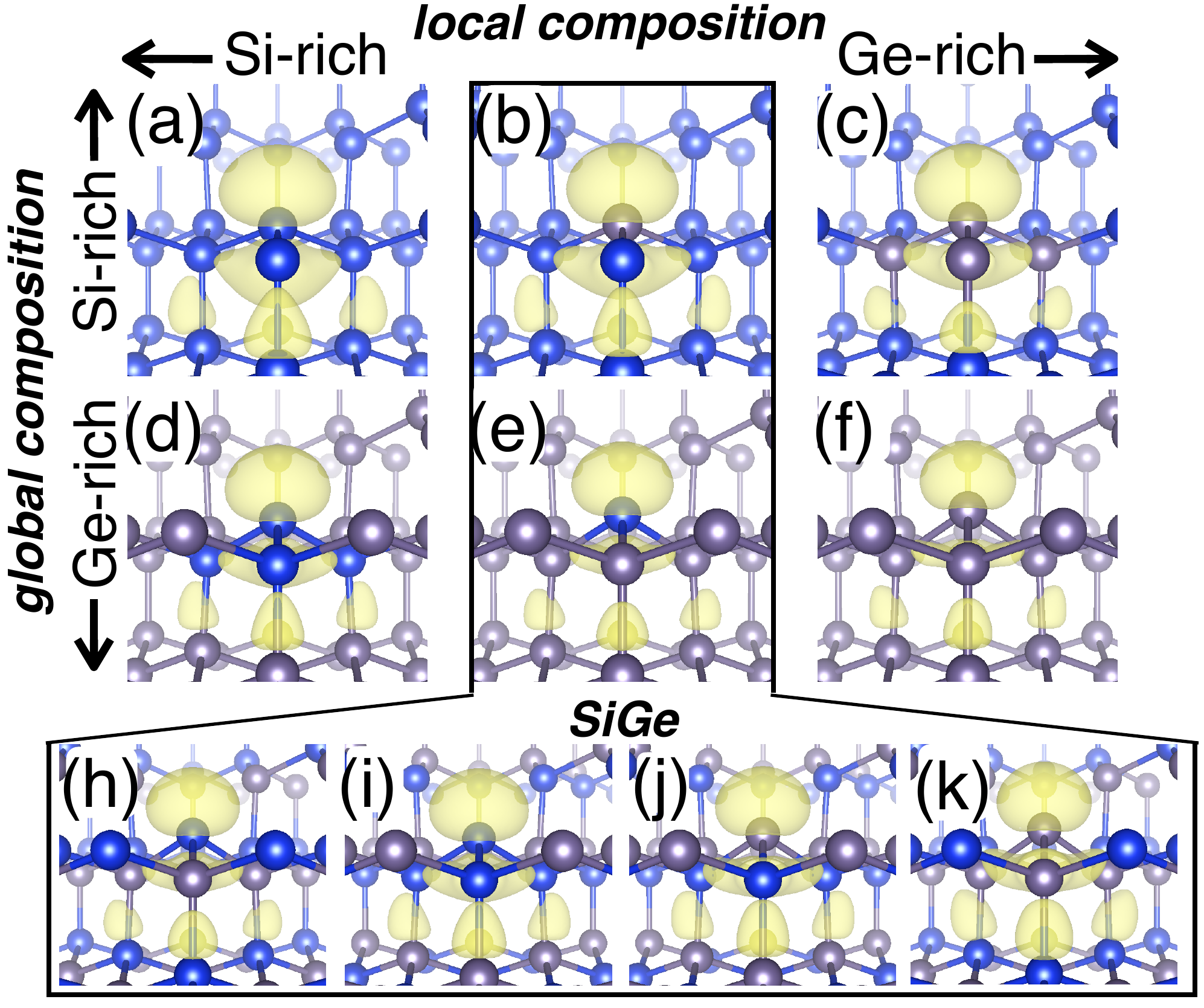}
\caption{Schematic of the investigated DBs for Si and Ge with variations in local and global environments, following the same atomic coloring and isosurface conventions as in Fig.~\ref{FIG:DIAGRAM}. 
DB structures showing different local bonding environments are illustrated in bulk Si for (a) $_{\rm Si}$Si$_{\rm Si}$, (b) $_{\rm Si}$Ge$_{\rm Si}$, and (c) $_{\rm Ge}$Ge$_{\rm Si}$ (Ge cluster). 
Analogous structures for bulk Ge include (d) $_{\rm Si}$Si$_{\rm Ge}$ (Si cluster), (e) $_{\rm Ge}$Si$_{\rm Ge}$, and (f) $_{\rm Ge}$Ge$_{\rm Ge}$.
Additional cases investigated in model 3C-SiGe Si$_{0.5}$Ge$_{0.5}$ alloy structures are shown for two types of Si DBs in (h) $_{\rm Ge}$Si$_{\rm Si}$ and (i) $_{\rm Si}$Si$_{\rm Ge}$, and Ge DBs in (j) $_{\rm Si}$Ge$_{\rm Ge}$ and (k) $_{\rm Ge}$Ge$_{\rm Si}$.
} \label{FIG:DBstructure}
\end{figure}

In a typical Si/\sige quantum well device architecture, a strained Si layer is sandwiched between \sige layers to introduce a two-dimensional electron gas (2DEG) confined in the strained Si layer as shown schematically in Figure~\ref{FIG:DIAGRAM}. 
The choice of \sige composition influences the details of the heterostructure band diagram, through both strain and band gap effects, with a typical design using $\sim$30\% Ge leading to high performing devices.\cite{Schaffler:1997f}  
The bottom layer is a relaxed \sige layer grown on a graded \sige{} virtual substrate that has compositional variations to tune both the strain in the quantum well and the resulting 2DEG quality while minimizing dislocations generated during the growth process.\cite{Schaffler:1997f}  
Even with optimized growth processes, threading dislocation densities  on the order of 10$^{5}-10^{6}$ cm$^{-2}$ have been reported.\cite{Mermoux.2010.10.1063/1.3272824,Schaffler:1997f}  
The top layers can also include relaxed \sige layers with additional doped layers to contribute carriers to the 2DEG. 
Owing to the presence of strain and likely compositional variations, including strain variations associated with defects and device patterning, we investigate the dependence of defect energy levels associated with DBs from Si and Ge, exploring the role of the local and global chemical environment. 
Specifically, we consider DBs from Si and Ge that are back-bonded to Si and Ge in bulk Si and Ge host materials to evaluate the scenarios of locally Si-rich and Ge-rich compositions, respectively. 
To avoid confusing notation with substitutional defects, we adopt a subscript in front to denote back-bonding that helps to clarify the local environment.
For example, an isolated Si DB in Si is redundantly labeled as $_{\rm Si}$Si$_{\rm Si}$, an isolated Ge$_{\rm Si}$ DB in Si is represented as $_{\rm Si}$Ge$_{\rm Si}$, while a locally-Ge rich region with all Ge nearest neighbors to the Ge$_{\rm Si}$ is labeled as $_{\rm Ge}$Ge$_{\rm Si}$.
We summarize schematics of studied DB structures in Fig.~\ref{FIG:DBstructure}.
We additionally consider the shifts in the band edges with composition and strain, relative to the defect levels, to give insight into how the electronic states associated with these defects may become localized within the band gap or act as sources of fixed charge for certain composition regimes. 
We find that for the strained Si typically adopted in \sige quantum well structures for qubits, Si and Ge dangling bond states exhibit  localized states within the band gap with low energy barriers for charge-state excitations that may contribute to charge-trapping and charge noise in these devices. These defects could likely be mitigated by a reduction in threading dislocation densities and improved DB passivation strategies. 

\section{Computational Methods}
All calculations were performed using density functional theory with the HSE06 screened hybrid functional and the projector-augmented wave (PAW) approach as implemented in the VASP code.\cite{Heyd:2003eg,*Heyd:2006dc, Kresse:1996kl,*Kresse:1996kg, Blochl:1994dx} 
This methodology yields accurate bulk properties such as band gaps, formation enthalpies, and band positions on an absolute energy scale, while simultaneously improving the description of charge-localization that is critical to accurately describing point defects such as vacancies.\cite{Varley:2012iw,Hinuma:2014fi,Peng:2013kx,Freysoldt:2014in,Varley:fx}
We considered a range of values for the mixing parameter, $\alpha$, of the screened Hartree-Fock contribution to the exchange energy from 25\% to 32\%; this range spans the values that yield accurate band gaps for Si and Ge. 
Band structures along the $\Gamma$--$X$ and $\Gamma$--$L$ directions in reciprocal space were computed including the effects of spin-orbit coupling to identify the predicted indirect band gaps for Si, Ge, and zinc-blende SiGe (50\% Ge); the values are summarized in Table~\ref{TAB:BULK}.
The calculated lattice parameters for Si ($a$=5.43~\AA) and Ge ($a$=5.65~\AA) are in excellent agreement with experimental values.\cite{madelung2004semiconductors}
Our calculated lattice constant for zinc-blende SiGe (as an ordered compound analogous to 3C-SiC) is 5.55~\AA, in good agreement with the 5.54~\AA\ measured for (disordered) 50\% \sige{} alloys at room temperature.\cite{Dismukes:1964iz}
For SiO$_2$, we considered the high-temperature tetragonal $\beta$-cristobalite phase as discussed by Coh and Vanderbilt,\cite{Coh:2008dt} with the lattice parameters optimized with $\alpha = 32\%$ included in Table~\ref{TAB:BULK}.

\begin{table}
\centering
\caption{\label{TAB:BULK}
Summary of the lattice constants and band gaps used for the bulk binary references in the diamond or zincblende lattices ($a$), and also the calculated $c$ axis for the biaxially strained Si to lattice match in in-plane lattice constant of Si$_{0.7}$Ge$_{0.3}$ ($a$=5.50). The $a$ and $c$ lattice parameters calculated for tetragonal $\beta$-cristobalite SiO$_2$ are also included. The band gap is direct for SiO$_2$ but indirect for Si, SiGe, and Ge.
The fraction of Hartree-Fock mixing used to calculate the properties for each material is defined in the text.
The low ($\epsilon_0$) and high-frequency ($\epsilon_\infty$) dielectric constants used in the finite size corrections for the thermodynamic and vertical charge-state transition levels are also included as taken from Ref.~\citenum{madelung2004semiconductors}.
}

\begin{tabular*}{\columnwidth}{@{\extracolsep{\fill}}ccccccccc}
\hline\hline
        Material & Si & Si (strained) & Ge & 3C-SiGe & SiO$_2$ \\
\hline  
$a$ (\AA) 	   	& 5.43 	& 5.50 & 5.68 	& 5.55 	& 5.05   \\ 
$c$ (\AA) 	   	& -- 		& 5.38 & -- 	& -- 		& 7.31   \\ 
$E_g$ (eV) 	& 1.16 & 0.94 & 0.81  & 1.20 	& 8.07 \\ 
$\Delta_{SO}$ & 0.05 	& 0.05 & 0.32 	&  0.19	& 0.01 \\
$E_{BP}$ (eV) 	& 0.17 	& 0.06 &  $-$0.20  & 0.05  & --  \\ 
$\epsilon_0$ ($\varepsilon_0$) 		& 12.1	& 12.1 & 16.2  & 13.95   & 4  \\
$\epsilon_\infty$  ($\varepsilon_0$)  & 12.1	& 12.1 & 16.2  & 13.95   & 2  \\ 
\hline\hline
\end{tabular*}
\end{table}

We employed model DB structures the same as detailed in Refs.\citenum{Weber:2007bk} and \citenum{Broqvist:2008jx}, where four atoms are removed and H is used to passivate the DBs generated in the removal of the atoms. 
We use 216-atom (reference) supercells, a planewave cutoff of 400~eV and a 2$\times$2$\times$2 grid of Monkhorst-Pack special $k$-points for all calculations and include spin-polarization for unpaired electrons. 
The fully-H-passivated structures were relaxed (residual forces less than 0.02~eV/\AA) initially, before one of the hydrogens was removed to create the DB; then, we selectively constrain a bulk-like region and further relax only the atoms associated with the DB and its nearest neighbors. 
We considered both first-nearest and second-nearest neighbors in the relaxations to evaluate the resulting structures and energies for the positive, neutral, and negative charge states of each DB. 
A schematic of the dangling bond configurations are shown in Fig.~\ref{FIG:DBstructure} and the insets of Fig.~\ref{FIG:DIAGRAM}. 
In addition to Si DBs in Si ($_{\rm Si}$Si$_{\rm Si}$), we consider the DBs of Ge$_{\rm Si}$ in Si ($_{\rm Si}$Ge$_{\rm Si}$), and Si$_{\rm Ge}$ in Ge ($_{\rm Ge}$Si$_{\rm Ge}$) as proxies for the influence of the local chemical environment for different \sige compositions. 
With the ordered SiGe structure (analogous to 3C-SiC), we evaluate both Si and Ge dangling bonds that are back-bonded to different sets of Si and Ge atoms: namely, Si DBs that are back-bonded to Ge ($_{\rm Ge}$Si$_{\rm Si}$) and Si (e.g. Si$_{\rm Ge}$ antisites, $_{\rm Si}$Si$_{\rm Ge}$), as well as Ge DBs back-bonded to Si ($_{\rm Si}$Ge$_{\rm Ge}$) and Ge (e.g. Ge$_{\rm Si}$ antisites, s$_{\rm Ge}$Ge$_{\rm Si}$). 
This provides another means to evaluate the variability of the energies of the states with local coordination.
An analogous procedure was performed for  $\beta$-cristobalite SiO$_2$ as a proxy for Si and Ge (Ge$_{\rm Si}$) DBs in oxide layers, also using 216-atom reference cells for the bulk SiO$_2$ lattice. For the oxide we only consider the Si and Ge$_{\rm Si}$ DBs for atoms that are back-bonded to O, and do not consider the additional effects of a more Ge-rich environment in next-nearest neighbor shells or beyond. 
For all defects, we characterize the degree of localization of any states induced within the band gap, i.e. deep versus shallow, from the density of states and projected charge densities.

\begin{figure}
\includegraphics[width=1\columnwidth]{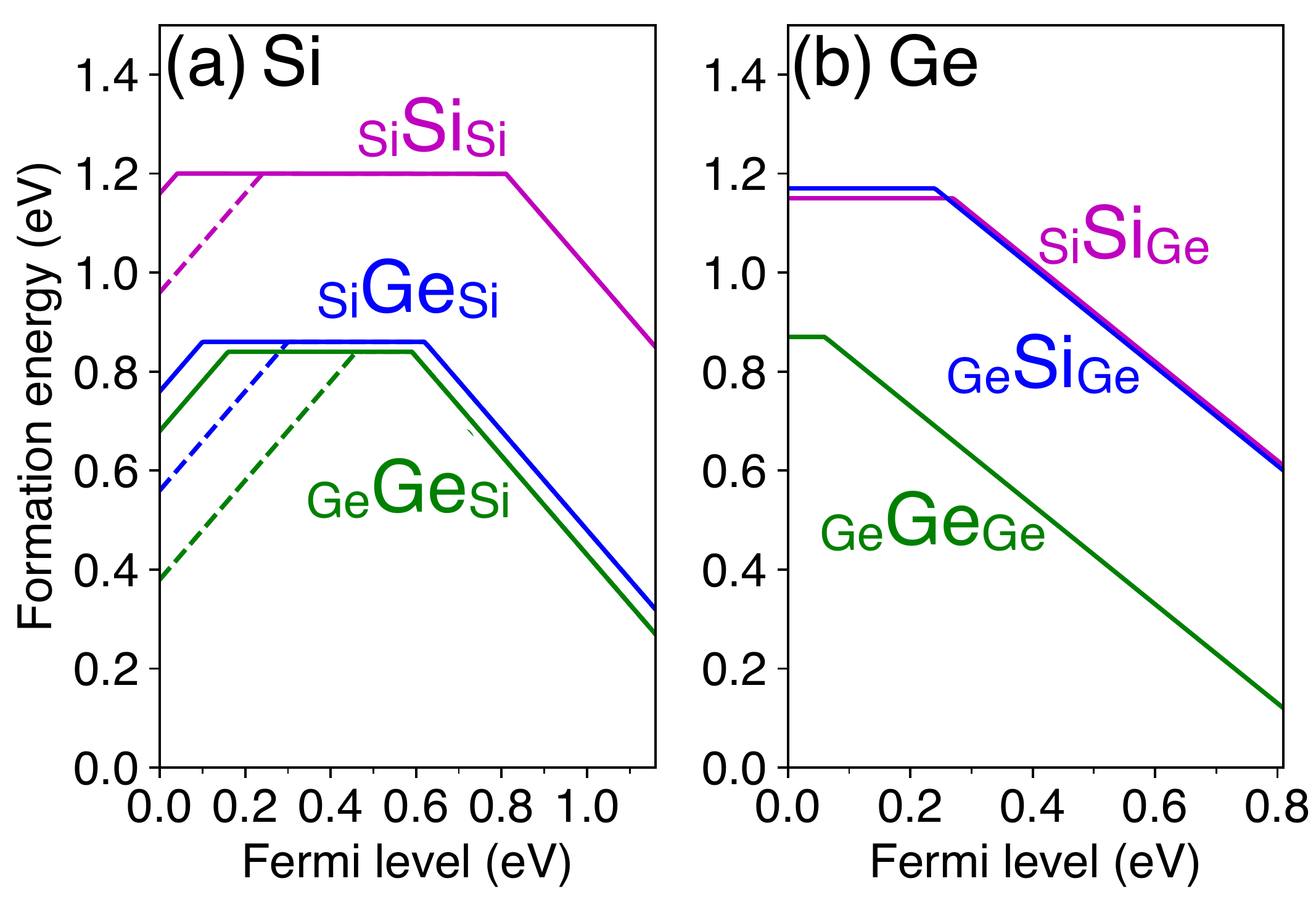}
\caption{Formation energies of Si and Ge DB defects with different charge states $+1$, $0$, and $-1$ (proportional to slope) are shown for the limiting cases of (a) bulk Si (a fully Si-rich matrix) and (b) bulk Ge (a fully Ge-rich matrix). 
The defect notation indicates with normal text the type of atom with the DB, while the leading subscript notation denotes the local environment around the atom with the DB, i.e., whether it is locally-bonded to Si or Ge nearest neighbors in the parent compound. 
The dashed lines in (a) for the +1 donor states include an additional stabilization correction of 0.2~eV, as described in the text. 
}
 \label{FIG:FORME}
\end{figure}

We calculate defect formation energies ($E^f$) and the associated charge-state transition levels ($\epsilon$) to determine the favorable charge states ($q$) of each defect under different conditions, using the standard supercell approach.\cite{VandeWalle:2004bk} 
For example, the formation energy of a generic DB defect in Si is given as a function of Fermi level ($\epsilon_F)$ by
\begin{align}\label{EQN-FORMATIONE}
&E^f[{\rm DB}^{q}](\epsilon_F) =  E_{\rm tot}[{\rm DB}^{q}] - E_{\rm tot}[ \textrm{DB:passivated}] + \mu_{\rm H} + q (\epsilon_F + \epsilon_v) + \Delta^q, 
\end{align}
where $E_{\rm tot}[{\rm DB}^{q}]$ represents the total energy of the supercell containing the defect $D$ in charge state $q$, and $E_{\rm tot}[{\rm DB: passivated}]$ represents the reference DB structure that has all DBs fully passivated. The chemical potential of the H atom removed from the passivated DB structure is included as $\mu_{\rm H}$, referenced as $\frac{1}{2}$ the energy of an isolated H$_2$ molecule.
The electrons are exchanged with a reservoir whose chemical potential is $\epsilon_{F}$, which we reference to the energy of the valence-band maximum (VBM), $\epsilon_v$. The effects of spin-orbit coupling were included as a correction to $\epsilon_v$ by including a shift of one-third the calculated spin-orbit splitting.
The $\Delta^q$ terms are additional corrections to account for finite-size effects resulting from the spurious periodic Coulomb interaction of charged defects and the alignment of the reference potentials in different charge states. 
The $\Delta^q$ were determined using the FNV scheme,\cite{Freysoldt:2009ih,*Freysoldt:2010gx} adopting static dielectric constants in Table~\ref{TAB:BULK}.
The FNV correction approach is the same as that adopted for Si DBs in Ref.\citenum{Chen:2017fi}, but differs from that of previous studies on the DBs in these systems that adopted either exclusively the potential alignment correction or the Makov-Payne approach to periodic image corrections,\cite{Weber:2007bk,Broqvist:2008jx} and can account for slight differences in the resulting values compared to these earlier studies.
A summary of the calculated formation energies of Si and Ge DBs in Si and Ge are shown in Fig.~\ref{FIG:FORME} and discussed later.

For the most stable defects, we report both the thermodynamic and optical (vertical), charge-state transition levels. The thermodynamic transition level, $\epsilon$, corresponds to the Fermi-level position at which a given defect changes its charge state in thermal equilibrium and depends on the formation energy difference obtained when each charge state assumes its relaxed configuration. 
Processes which can change the charge state of the defect on a  timescale faster than atomic relaxations ($\sim$10$^{-13}$ s), i.e. optical excitations and tunneling processes, are described by taking the difference of formation energies for two charge states with the structure frozen in the configuration of the initial charge state.
We denote the former, thermodynamic transition levels as $\epsilon(q/q^\prime)$ and the latter as $\epsilon^o(q/q^\prime)$ for transitions between an initial charge state $q$ and final charge state $q^\prime$. 
Additional finite-size corrections for vertical transition energies were determined using the approach detailed by Gake {\em et al.} in Ref.~\citenum{Gake:2020:10.1103/physrevb.101.020102}, using the high-frequency dielectric constants in Table~\ref{TAB:BULK}. 

For the calculation of the $\epsilon^o$ values, we use a configuration coordinate diagram analysis to report optical absorption and emission energies according to the Franck-Condon principle. These energies are also relevant to understanding non-radiative recombination processes.\cite{Alkauskas:2014kk,Alkauskas:2016kf}  
Following this principle, the onset energy for optical transitions corresponds to the zero-phonon line (ZPL), which is obtained from the energy difference between the relevant thermodynamic charge state transition level and the band edge involved in the excitation process. The relaxation energies (Franck-Condon shifts, $R^q_{\rm FC}$) are obtained by single-shot calculations of one charge state in the relaxed configuration of the other charge-state involved in the excitation process, which leads to shifts in the thermodynamic transition levels according to the ZPL and $R^q_{\rm FC}$ values. 
We additionally evaluate the classical energy barrier ($\Delta E_b$) for a transition from the excited to ground states by the energy difference between the intersection of the ground and excited state potential energy curves and the minimum energy of the excited state.\cite{Alkauskas:2014kk,Alkauskas:2016kf}  
The magnitude of this barrier influences the transition rates between ground and excited charge states which may contribute to charge noise in the qubits at different frequencies, with the sum of $\Delta E_b$ and the associated thermodynamic transition levels giving a good estimate of experimental signatures of the defect levels such as those measured with deep-level transient spectroscopy.\cite{Wickramaratne:2018:10.1063/1.5047808}

Lastly, we consider the full composition space of Si$_{1-x}$ Ge$_x$ (random) alloys using relaxed 64-atom diamond cubic model structures  constructed from two interpenetrating face-centered cubic ($fcc$) sub-lattices of 32-atom special quasi-random structures (SQS), as described in Ref.\citenum{vonPezold:2010cr} and previously adopted for zinc-blende lattices in Refs.\citenum{Varley:CdZnOSthermo} and \citenum{Varley:2017hv}.
This procedure approximates the true randomness of the diamond lattice by assuming the correlations between the two identical $fcc$ sub-lattices are independent, which need not  be exactly the case.
We generate 15 alloy structures for compositions interpolated between pure Si to ordered SiGe, and 15 structures for compositions between SiGe and Ge. This leads to a total of 33 structures spanning the full \sige composition range, where the lattice constants were linearly interpolated between the compositions from the calculated lattice parameters in Table ~\ref{TAB:BULK}. 
To evaluate the properties of the strained Si along the (001) direction, we fixed the in-plane lattice constants to that of Si$_{0.7}$Ge$_{0.3}$ (calculated to be $a$=5.50~\AA) and relaxed the perpendicular lattice constant, as given in Table~\ref{TAB:BULK}.
All alloy calculations adopted a plane-wave cutoff of 400~eV and a 2$\times$2$\times$2 $\Gamma$-centered $k$-point grid and additionally sampled reciprocal space points in the vicinity of the $X$ and $L$ points to more accurately evaluate the conduction band extrema of the relaxed alloy structures. Spin-orbit effects were evaluated with a single-shot calculation from the final relaxed structures.

To evaluate the band offsets as a function of stoichiometry in the \sige SQS alloy structures, we employed band alignments based on the branch-point energy ($E_{BP}$) evaluated for each composition, due to the difficulty and ambiguity of constructing explicit strained interface or surface models for a more direct approach.\cite{Varley:2017hv,Varley:fx}
The $E_{BP}$ represents an effective energy (e.g. a charge-neutrality level) where the bulk states change from predominantly valence-band-like (donor-like) character to conduction-band-like (acceptor-like) character and has been very successful in defining a reference level from which to align the unstrained band edges of tetrahedrally-coordinated semiconductors. \cite{Tersoff:1986vx,Tejedor:2001do,Monch:1996kk,VandeWalle:2003tg,Robertson:2013ha,Schleife:2009he,Cardona:1987fl,Hinuma:2014fi} 
Here we evaluate $E_{BP}$ from the calculated eigenvalues for the alloy structures following the procedure detailed in Refs.\citenum{Schleife:2009he} and \citenum{Varley:2017hv} using a $k$-point weighted average over 64 valence bands and 32 conduction bands in the alloy structures. (The number of bands included in the weighted averages are scaled from the 2 valence bands and 1 conduction band of the primitive unit cells of diamond and zinc-blende lattices justified in Ref.\citenum{Varley:fx} and the references therein.)
We note that there is a large sensitivity of this approach to the particular number of bands chosen for cases larger than the 2-atom primitive unit cells, as has been investigated via a variety of approaches including explicit band offset calculations, Green's function approaches, and defect calculations.\cite{Tersoff:1986vx,Tejedor:2001do,Monch:1996kk,VandeWalle:2003tg,Cardona:1987fl,Hinuma:2014fi,Varley:fx} Therefore, we use the $E_{BP}$ as a simple, but consistent metric to evaluate how the alloy valence band position shifts relative to pure Si and Ge; indeed, our results indicate the relationship is effectively linear with composition between Si and Ge (i.e., negligible valence band bowing).

\section{Results and Discussion}


We begin by summarizing the electronic structure of \sige that hosts the studied DB defects. In Fig.~\ref{FIG:DB1} we illustrate the computed band gaps and band edges over the entire composition range for the \sige alloys, as aligned via the $E_{BP}$ in the SQS supercells. 
From this approach we calculate a natural valence band offset (VBO) of 0.37~eV between pure Si and Ge, which is in good agreement with a value of 0.38~eV reported for recent calculations adopting hybrid functionals with many-body perturbation theory corrections.\cite{Hinuma:2014fi}
We also are able to capture the transition of the indirect band gap from occurring at the $\Delta$ point (near an $X$-valley) to an $L$-valley, which our results show occurs for increasing Si content at $\sim$25\%. This composition is slightly higher than reported values that suggest the transition occurs closer to 15\% Si.\cite{Anonymous:2p4fYeLj,Braunstein:1958wd}. This agreement is good, considering the slight overestimation of the Ge band gap that tends to overestimate the Ge fraction for the transition and the limited number of approximate random alloy models that we probed in the study. The results in Fig.~\ref{FIG:DB1} also suggest a virtually linear change in the unstrained valence band position with composition and provides a baseline from which to interpret the positions of the defect levels with $\sim$0.1~eV precision.


\begin{figure} \label{FIG:DB1} 
\includegraphics[width=0.99\textwidth]{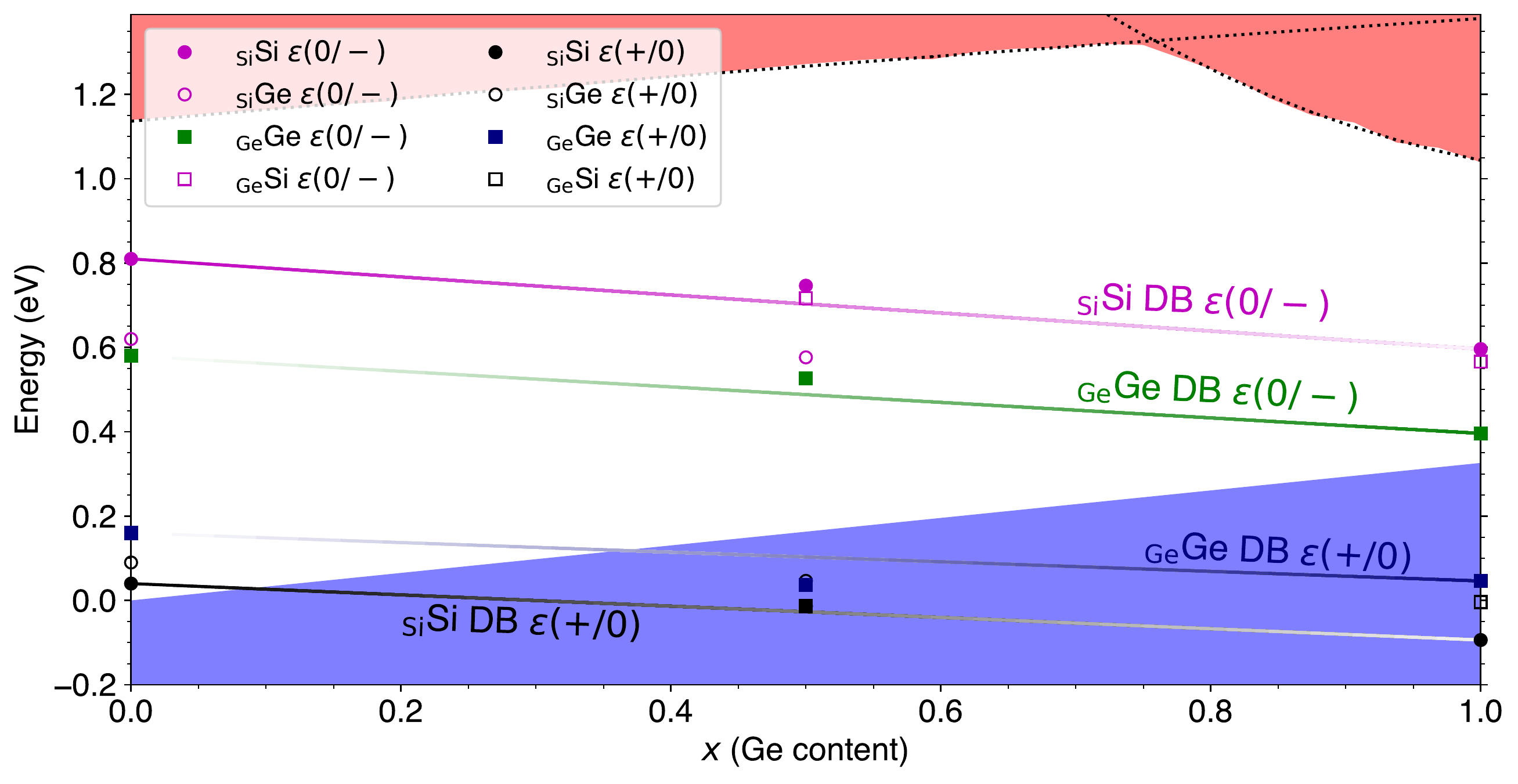} 
\caption{ The computed $\epsilon(+/0)$ and $\epsilon(0/-)$ transition levels of Si and Ge DBs in Si$_{1-x}$Ge$_x$, plotted as a function of composition and shown relative to the computed band edges. 
The valence band is shown shaded in blue, while the conduction band is shaded in red, and are aligned on an absolute energy scale as described in the text. 
The dotted lines show the composition dependence of the $\Delta$ and $L$ conduction band minima (CBM), which cross at $x \sim 0.75$ and comprise the CBM at different compositions.
The DB defect levels are interpolated between the values for Si back-bonded Si DBs (solid circles) and Ge back-bonded Si (open squares) spanning from bulk Si ($_{\rm Si}$Si$_{\rm Si}$) to bulk Ge ($_{\rm Si}$Si$_{\rm Ge}$), and for Ge back-bonded Ge DBs (solid squares) and Si back-bonded Ge (open circles) spanning from bulk Si ($_{\rm Ge}$Ge$_{\rm Si}$) to bulk Ge ($_{\rm Ge}$Ge$_{\rm Ge}$).  
Shaded regions correspond to the spread associated with the different back-bonding environment within a given host composition.
Explicitly calculated values for the 50\% alloy (3C-SiGe) are shown for these cases, which correspond to DBs on antisite defects of Si$_{\rm Ge}$ ($_{\rm Si}$Si$_{\rm Ge}$) and Ge$_{\rm Si}$ ($_{\rm Ge}$Ge$_{\rm Si}$), respectively.  
The calculated values indicate their sensitivity to the local chemical environment (see Fig.~\ref{FIG:DBstructure}), with the symbols as summarized in the legend all within 0.1~eV of the respective interpolated values from the pure compositions. 
}
\label{FIG:DB1}
\end{figure}

\subsection{Dangling bonds in \sige alloys}

The calculated formation energies of the various DBs in Si and Ge are summarized in Fig.~\ref{FIG:FORME}, which highlights the  overall influence of local chemical environment that will be discussed further below. 
Note that we show results for both Si and Ge DBs in both bulk Si (Si-rich) and bulk Ge (Ge-rich) matrices, to represent the 6 different general types of DBs in the alloys, here extrapolated to the dilute limits at the ends of the concentration range. 
Specifically, these are the three types of local bonding environments within two different limiting host electronic structures. 
While it is difficult to  predict directly a (thermodynamic) concentration of DB-type defects from the calculated $E^f$ of the model structures via Eqn.~\ref{EQN-FORMATIONE}, the calculated charge-state transition levels are quantitatively meaningful and qualitative concetration trends still hold.
In particular, Fig.~\ref{FIG:FORME} shows that DBs in both Si and Ge have low formation energies, suggesting their prevalence, with Ge DBs found to be more thermodynamically favorable by $\sim$0.35~eV than Si DBs. 
Interestingly, the $E^f$ for the neutral DBs are largely independent of the host material and more sensitive to the local environment, with more Ge-rich local environments leading to greater stability (lower $E^f$). 
Overall, these results suggest that DB-type defects can easily form throughout the full composition of \sige and may be present in non-negligible concentrations in typical Si/\sige quantum dot qubit architectures, particularly in the vicinity of surfaces or interfaces where strain and compositional variations may influence the bonding environment.
They also suggests that more Ge-rich regions may be more susceptible to the formation of greater concentrations of DB defects. 
In the following sections we discuss the different types of DBs in more detail, focusing on the electronic consequences of these defects and their possible roles in charge noise resulting from carrier capture and emission.

\subsubsection{Pure silicon}
For a Si DB in Si (i.e., Si back-bonded to Si, notated ``$_{\rm Si}$Si$_{\rm Si}$''), we calculate that both the $\epsilon(+/0)$ and $\epsilon(0/-)$ transition levels fall within the band gap, at 0.04 and 0.81~eV above the unstrained VBM, respectively, as shown in Fig.~\ref{FIG:FORME}(a). 
These values are summarized in Table~\ref{TAB:EPS} and correspond to the left edge in \fig{FIG:DB1}, which plots the DB transition levels across the \sige composition range (discussed later).
The deeper $\epsilon(0/-)$ level agrees well with experimentally reported values of 0.8-0.84~eV\cite{Poindexter:1984ky, Gerardi:1986dh,Johnson:1983eq} and previous calculations using various hybrid functionals and the $G_0W_0$ approximation of 0.7-0.8~eV.\cite{Broqvist:2008jx,Chen:2017fi} 
The agreement is slightly worse for the $\epsilon(+/0)$, which was previously calculated to be between 0.1-0.3~eV and experimentally measured to be 0.26-0.31~eV from a combination of capacitance-voltage ($C$-$V$) and electron paramagnetic resonance (EPR) measurements for the donor transition in thermally oxidized Si.\cite{Poindexter:1984ky, Gerardi:1986dh,Johnson:1983eq,Broqvist:2008jx,Chen:2017fi} 
The theoretical reports for this value are highly sensitivity to details of the calculations; for example, the previous hybrid and HSE+$G_0W_0$ results show a notable sensitivity in the position of the $\epsilon(+/0)$ level with the calculated valence band position and the self-interaction error was found to be larger for the donor state based on the magnitude of the quasiparticle corrections to the transition level (0.2~eV). \cite{Broqvist:2008jx,Chen:2017fi} 
We also find a larger sensitivity of this transition level with supercell size, finding a deviation of 0.1~eV between the finite-size corrected $\epsilon$(+/0) values for 64- and 216-atom supercells, whereas the $\epsilon(0/-)$ was already converged within 0.03~eV for 64-atom supercells.
To account for this large technical uncertainty and in light of the experimental measurements, in our analysis that follows we consider the energies both as-calculated with our HSE06 approach and also consider scenarios with the (+/0) levels shifted 0.2~eV deeper (higher) within the band gap, and thus evaluate the sensitivity of our conclusions to this uncertainty. 


Next, we consider Ge DBs in Si, where a substitutional Ge atom is back-bonded to 3 Si atoms (notated ``$_{\rm Si}$Ge$_{\rm Si}$'' in the figures).
We also consider Ge back-bonded to 3 Ge in a Si matrix (notated ``$_{\rm Ge}$Ge$_{\rm Ge}$''), which we utilize as a proxy to study the range of influence of the local chemical environment in the alloy, here as a Ge ``cluster.''
The relative local atomic relaxations of the isolated Ge and Si DBs are largely similar, where the neighboring bonds relax slightly inward for the donor state ($-$1-2\%), remain nearly unchanged for the neutral state, and relax slightly outward for  the acceptor state (+2-3\%). 
However, we find that the levels shift toward midgap for both the $\epsilon(+/0)$ and $\epsilon(0/-)$ transitions. Relative to the Si DB levels, we find Ge back-bonded to 3 Si (3 Ge) leads to an upward shift of $\epsilon(+/0)$ by 0.05 (0.12)~eV and a downward shift of $\epsilon(0/-)$ by 0.19 (0.23)~eV, as shown in \fig{FIG:FORME} and summarized in Table~\ref{TAB:EPS}. 
These results imply that Ge DBs in the limit of more Si-rich \sige are more problematic than native Si DBs from the standpoint of Shockley-Read-Hall (SRH) recombination, for which non-radiative recombination rates increase exponentially as the levels approach the middle of the band gap. This is shown schematically in Fig.~\ref{FIG:DB1} by the proximity of the Ge DB transition levels (e.g., green dashed line) to the solid grey line that represents the midgap energy as a function of composition, compared to the Si DBs (e.g., magenta dashed line).
These results further suggest that DB-related capture and emission of free carriers, and the associated changes in spatially-localized charges, may become a more significant source of charge noise in regions containing higher local Ge content for Si-rich material.

The nominally undoped, intrinsic Si layers in typical Si/\sige quantum dots can be expected to have an equilibrium Fermi level close to mid-gap. Based on the positions of the DB levels, Si DBs would preferentially adopt neutral charge states over a wide range of Fermi level and would be unlikely to contribute to charge noise in the absence of other charge excitation events (e.g., optical excitation, discussed later). Ge DBs in Si exhibit $\epsilon(0/-)$ transition levels very close to mid-gap that could make these DBs preferentially adopt negatively-charged configurations or fluctuate between charge states. Slight deviations in the Fermi level, such as through unintentional impurities or band bending in the device, and also variations from applied gate voltages, may also influence the preference between neutral and acceptor configurations of the DBs in Si. Thus, there is the possibility that dynamic changes in the preferential charge state of any Ge DBs that may be present in Si layers could contribute to charge noise, depending on the rates of charge transitions, which will be considered later.

\begin{table*}
\centering
\caption{\label{TAB:EPS}
Summary of the calculated thermodynamic transition levels, reported relative to the VBMs of each respective material. The Si DB and Ge DB refer to the levels of an isolated Si or Ge DB back-bonded to the host matrix (e.g.  $_{\rm Si}$Si$_{\rm Si}$ and  $_{\rm Ge}$Si$_{\rm Ge}$ for Si in bulk Si and bulk Ge, respectively). 
Cluster DBs refer to atoms that are locally coordinated to the same type within another matrix, e.g. Si coordinated to 3 other Si in a Ge or SiGe matrix ($_{\rm Si}$Si
$_{\rm Ge}$) as described in the text. 
All values are reported in~eV.
}
\begin{tabular*}{\columnwidth}{@{\extracolsep{\fill}}rccrrc}
\hline\hline
       			 & Si & Si (strained) & Ge & 3C-SiGe & SiO$_2$ \\
\hline  
Si DB    		&  &  & &  	&    \\ 
$\epsilon(+/0)$ 	& 0.04 	& $-$0.03 &   $-$0.33& -- 	& 3.57  \\ 
$\epsilon(0/-)$	& 0.81	& 0.80 &  	0.24	& -- 	& 6.25  \\ 
$\epsilon(+/-)$	& 0.43 	& 0.37 &  $-$0.04	& -- 	&  4.91 \\ 
Ge DB    		&  &  & &  	&    \\ 
$\epsilon(+/0)$ 	& 0.09	& 0.02 & $-$0.28	& -- & 2.37  \\ 
$\epsilon(0/-)$	& 0.62	& 0.63 &  0.07	& -- 	& 4.56  \\ 
$\epsilon(+/-)$	& 0.36	& 0.32  &  $-$0.11	& -- 		& 3.46  \\ 

Si cluster DB    	&  &  & &  	&    \\ 
$\epsilon(+/0)$ 	& --	& -- & $-$0.42	& $-$0.23	& --  \\ 
$\epsilon(0/-)$	& --	&--  &  0.27	& 0.65	&  -- \\ 
$\epsilon(+/-)$	--& --	& --  &  $-$0.07	& 0.21	& --   \\ 
Ge cluster DB   &  &  & &  	&    \\ 
$\epsilon(+/0)$ 	& 0.16	& -- &  -- 	& $-$0.16	& --   \\ 
$\epsilon(0/-)$	& 0.58	& -- &  -- 	& 	0.46 & --   \\ 
$\epsilon(+/-)$	& 0.37	& --  &  --	& 0.15	& --   \\ 

\hline\hline
\end{tabular*}
\end{table*}

\subsubsection{Pure germanium and \sige alloys}
Now we turn to the pure Ge extreme of the composition range, again considering both Ge and Si (dilute impurity, notated ``$_{\rm Ge}$Si$_{\rm Ge}$'') DBs.
The results are summarized in Table~\ref{TAB:EPS} and correspond to the right edge of Fig.~\ref{FIG:DB1}.
A key difference between the DBs in Ge compared to Si is that, in Ge, they are only found to adopt neutral or acceptor configurations [i.e., (0/--) transitions], but not also donor states [i.e., (+/0) transitions] as in Si (see Fig.~\ref{FIG:FORME}). 
From Fig.~\ref{FIG:DB1}, we can see the origin of this effect arises from the higher VBM of Ge, as well as the lower transition energies, which pushes the donor transitions below the VBM and outside the band gap for high Ge concentrations.  
These results are consistent with previous studies on Ge DBs that also find the transitions from the donor state are either resonant with the valence band or within the band gap but very close to the VBM.\cite{Weber:2007bk,Broqvist:2008jx}
We predict the transition for these donor transitions to be pushed below the VBM occurs at a Ge concentration between $\sim$25 and 75\%, depending on whether we use the as-calculated or empirically corrected values, as depicted in \fig{FIG:DB1} and discussed further below.

Our results indicate that the $_{\rm Ge}$Si$_{\rm Ge}$ DB has a slightly deeper acceptor level, falling 0.24~eV above the VBM for pure Ge, compared to the $_{\rm Ge}$Ge$_{\rm Ge}$ DB, which has the $\epsilon(0/-)$ transition 0.07~eV above the VBM. 
We find that for the $_{\rm Si}$Si$_{\rm Ge}$ DB, the acceptor level falls at virtually the same energy (0.27~eV above the VBM) as for the $_{\rm Ge}$Si$_{\rm Ge}$ DB.
Comparing the relative differences of the energy levels of the Si and Ge DBs ($\sim$0.2~eV differences), along with the variations of the DB levels with the different back-bonding chemical environments ($\sim$0.05~eV), suggests that the local environment of the DB is a relatively minor perturbation on the defect level position that leads to a relatively narrow, but non-negligible, distribution of defect energy levels with local compositional variations (e.g., local clustering of Ge or Si in dilute alloys, or local ordering in more concentrated alloys).

Using the results for the pure matrix materials, we have interpolated the energy levels of the Si and Ge DBs across the entire composition range of \sige in Fig.~\ref{FIG:DB1}, where the thermodynamic transition levels are plotted as the various dashed lines. 
In addition to the analysis of local chemical variations on these levels presented above for the pure Si and Ge matrices, we further evaluated the DB energies explicitly for the zinc-blende ordered 50\% Ge 3C-SiGe phase (symbols in \fig{FIG:DB1}); we also use these calculations to assess how well the linear interpolation between Si and Ge holds for the DB levels. 
We considered the DBs associated with ``normal" Si and Ge (which are back-bonded to Ge, and Si respectively in the 50\% ordered alloy), as well as for antisite defects of Si$_{\rm Ge}$ and Ge$_{\rm Si}$ to sample the possible local environments in 3C-SiGe. 
The results are for all cases are included in Fig.~\ref{FIG:DB1}, where we find that the Ge DB levels are slightly more sensitive to the local environment than the Si DBs, but the overall magnitude is found to be $\sim$0.05~eV. 
Both the Si and Ge DBs for 3C-SiGe are calculated to within 0.1~eV of the average of the respective DBs in pure Si or Ge, indicating that the interpolation is justified for both the donor and acceptor transition levels and also further supporting the conclusion that local compositional variations can lead to about 0.1~eV variation of the levels.
These energies are very similar to the reported shifts in the transition levels going from isolated to ``cluster" DBs summarized in Table~\ref{TAB:EPS} for the Si and Ge limits. Therefore both the local environment and global composition-dependent electronic structure contribute to position of any DB-related defect levels, with the host Si$_x$Ge$_{1-x}$ composition and dominant chemical species (e.g. a Ge DB or Si DB) playing the largest role in their energy distribution. We note that the local environment {\em can} have a more pronounced impact on the energy landscape governing charge excitation events, as we detail later when discussing the configuration coordinate diagrams.

\subsubsection{Dangling bonds in strained layers} 

The active region in a Si/\sige quantum dot device arises from the 2DEG formed in the biaxially strained Si (001) layer that forms the quantum well, at the interface with the SiGe. As stated previously, we considered the scenario where the the in-plane lattice constant of Si was set to that calculated for Si$_{0.7}$Ge$_{0.3}$ (5.50~\AA) and relaxed in the out-of-plane direction, as summarized in Table ~\ref{TAB:BULK}. This strain leads to shifts of the Si band edges, in addition to possible shifts of the DB levels associated with the altered local symmetry from the distortion of the diamond lattice. 
For this case of 1.3\% biaxial tensile strain, we calculated an upward shift in the Si valence band edge of 0.11~eV and an overall narrowing of the band gap by 0.22~eV, which is consistent with previous calculations.\cite{Anonymous:2p4fYeLj}
This was determined from the relative strain in the in-plane and out-of-plane directions and the deformation potential $b$ as detailed in Ref.~\citenum{Anonymous:2p4fYeLj}, with our calculated value of $b=-2.45$~eV in good agreement with the reported theoretical ($-$2.35~eV) and experimental ($-$2.1$\pm0.1$~eV) values.  
From the predicted band gap and band edge shifts shown in Fig.~\ref{FIG:DB1}, we find that Si$_{0.7}$Ge$_{0.3}$ has a small change in the conduction band position relative to unstrained Si ($+0.07$~eV), with the decrease in band gap resulting from a larger relative upward shift of the valence band. Combining these values gives us a calculated conduction band offset of the strained Si relative to the relaxed Si$_{0.7}$Ge$_{0.3}$ of $-$0.18~eV, in good agreement with previously reported values around $-$0.21~eV and $-0.15$~eV.\cite{Virgilio.2006.10.1088/0953-8984/18/3/018, Schaffler:1997f}

For the DB states in strained Si, we find that their relative energies do not significantly change compared to unstrained Si. 
For example, the calculated $\epsilon(0/-)$ transitions for Si and Ge DBs appear 0.80~eV and 0.63~eV above the strained Si VBM, respectively.
The calculated Si and Ge DB  transitions for $\epsilon(+/0)$ shift the most from unstrained values (lower by  $\sim$0.07~eV, see Table~\ref{TAB:EPS}), which likely stems from the larger influence of the strain on the valence band and the previously mentioned errors associated with the theoretical description of the $\epsilon(+/0)$.
Thus, despite the perturbed band edges and reduction in the band gap, the population of DBs in various charge states in strained Si layers are expected to be similar to those in unstrained layers, with a sensitivity of the relative populations of charged DBs most sensitive to the Fermi level. \cite{Poindexter:1984ky,Broqvist:2008jx} 

\begin{figure}
\includegraphics[width=1.0\textwidth]{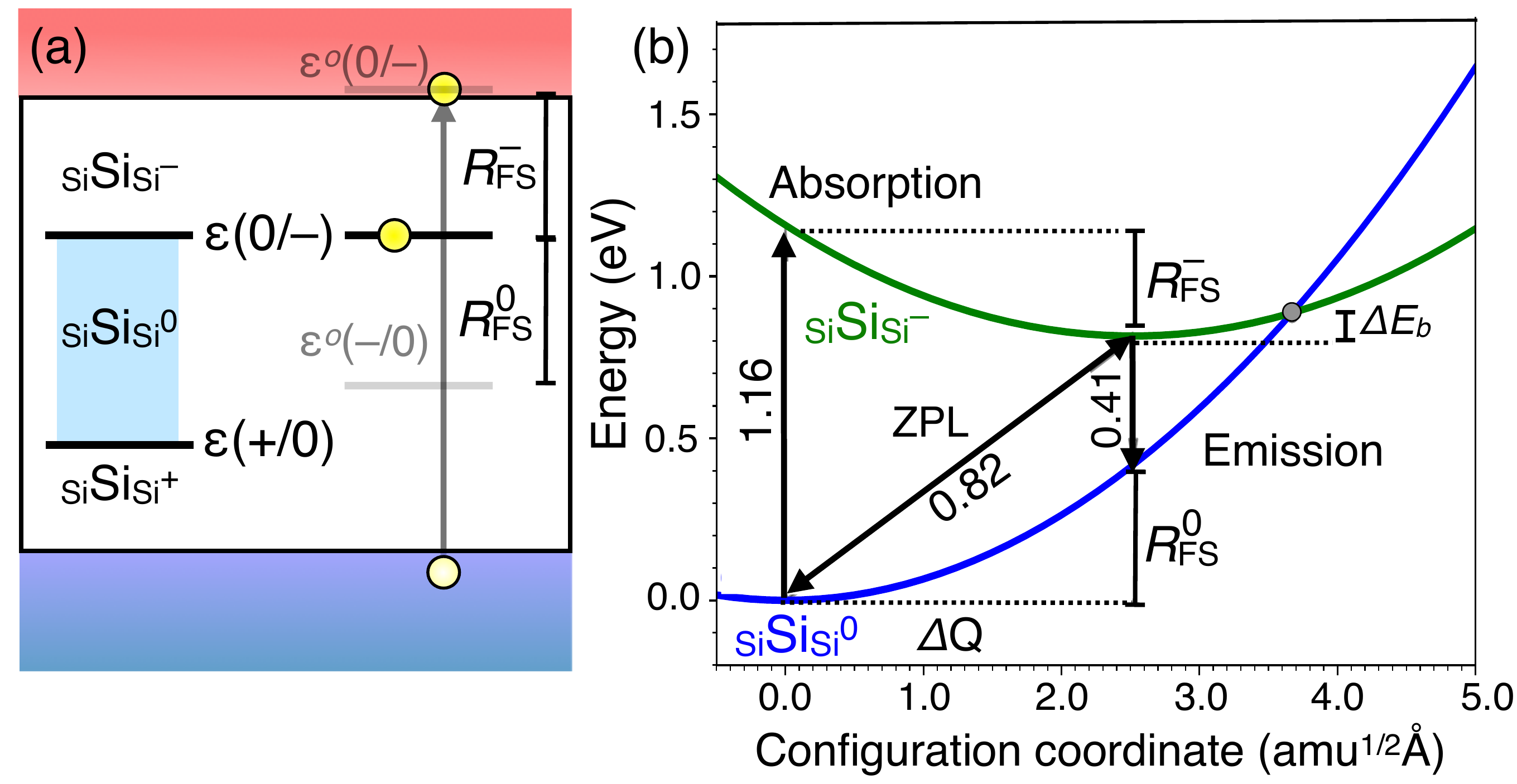}
\caption{(a) Schematic of the defect level positions associated with the Si DB in Si  ($_{\rm Si}$Si$_{\rm Si}$) and an example excitation of electron capture from the valence band at a neutral Si DB. The corresponding configuration coordinate diagram is shown in (b), illustrating the relevant parameters described in the text. The illustrated excitation process shows how the thermodynamic ($\epsilon$) and vertical ($\epsilon^0$) transition levels are related to the zero-phonon line (ZPL), the Franck-Condon shifts ($R^q_{\rm FS}$) and associated absorption and emission energies associated with the defect.
} \label{FIG:CCD0}
\end{figure}

\begin{figure}
\includegraphics[width=1\columnwidth]{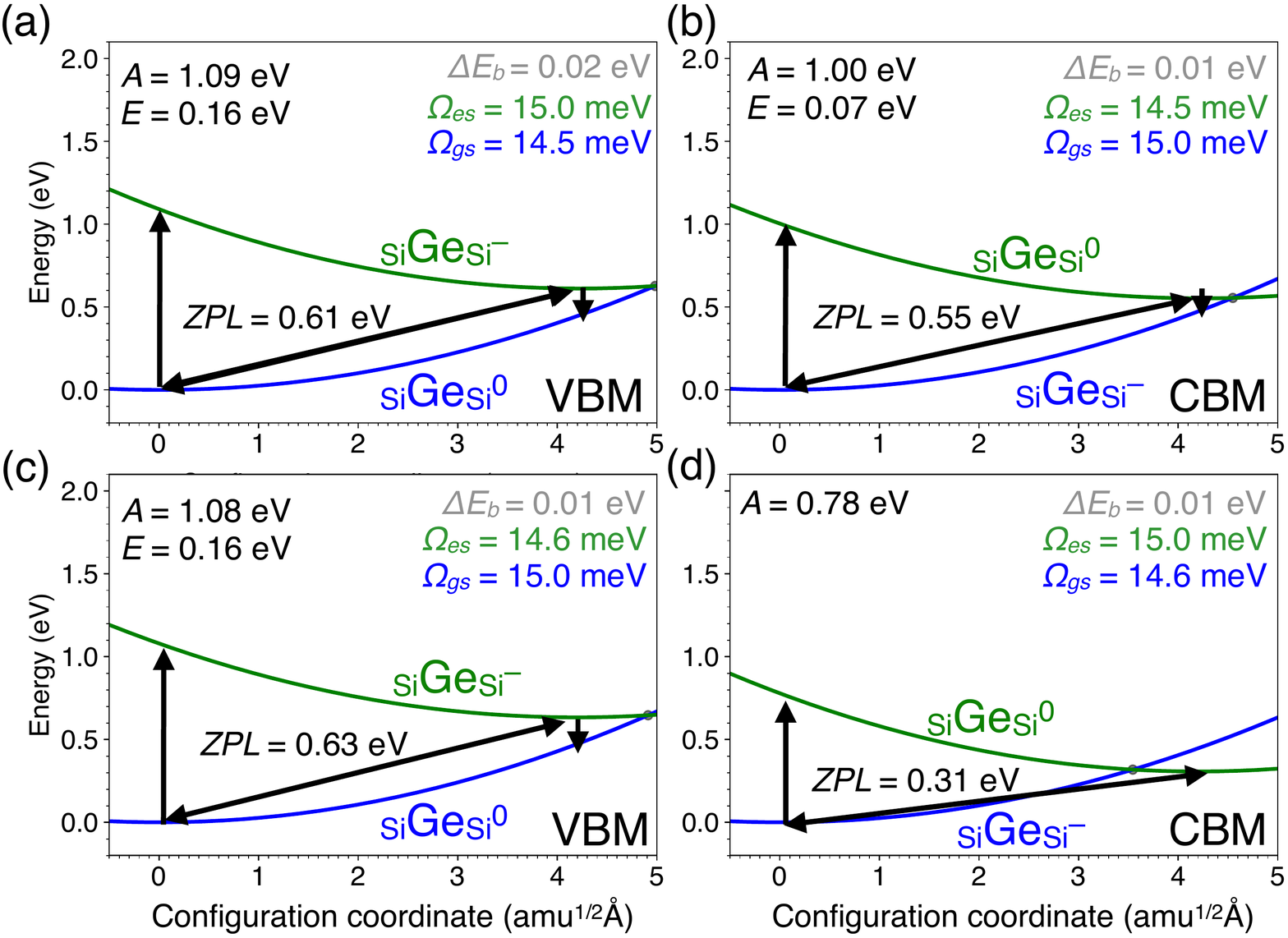}
\caption{Summary of the calculated configuration coordinate diagrams denoting vertical excitation energies (carrier absorption and emission) for a Si back-bonded Ge DB  ($_{\rm Si}$Ge$_{\rm Si}$) in unstrained (a and b) and strained (c and d) Si. The ZPL denotes the position of the thermodynamic transition level relative to the VBM or CBM (summarized in Table ~\ref{TAB:EPS}). The diagrams also include the vibrational energy associated with the ground ($\Omega_{gs}$) and excited states ($\Omega_{es}$), and the classical barrier height ($\Delta E_b$) determined from the crossing points of the potential energy surfaces as seen in Fig.~\ref{FIG:CCD0}b. 
The one-dimensional configuration coordinate relates to the bond lengths of the DB atom and its nearest neighbors. 
} \label{FIG:CCD}
\end{figure}

\subsubsection{Configuration coordinate diagrams for the dangling bonds}
To elucidate the energetics of charge excitation events involving the dangling bonds, we use a pseudo-one dimensional configuration coordinate diagram (CCD) analysis. The CCD analysis allows us to readily assess the energetics of charge transfer events involving electrons or holes emitted or captured by a localized DB state at some average rate, leading to charge fluctuation events that may contribute to charge noise at characteristic frequency ranges.
A schematic example illustrating electron capture at at neutral Si DB in Si is shown in Fig.~\ref{FIG:CCD0}, which also illustrates the associated CCD and relevant parameters for understanding and characterizing possible defect excitation processes, described below.
We summarize the calculated CCDs for Si and $_{\rm Ge}$Ge$_{\rm Si}$ DBs in Si in Fig.~\ref{FIG:CCD}, highlighting transitions between the neutral and negatively-charged acceptor configurations that can exchange carriers with either the valence or conduction band edges. 

For a given excitation process, the CCD includes information about the zero-phonon line (ZPL), optical absorption ($A$) and emission  ($E$) energies related to the ZPL through Franck-Condon shifts ($R^q_{\rm FS}$), and a classical barrier ($\Delta E_b$) associated with the crossing of the excited and ground state potential energy surface (see Fig.~\ref{FIG:CCD0}).
Within these pseudo-one dimensional CCDs (the configurational variations are three-dimensional, but the analysis is mapped to a single coordinate), 
as in Figures~\ref{FIG:CCD0} and \ref{FIG:CCD}, the ZPL corresponds to the position of the $\epsilon(0/-)$ thermodynamic transition level relative to the relevant band edge with which carriers are being exchanged.
We also include the calculated vibrational energies of the ground ($\Omega_{gs}$) and excited ($\Omega_{es}$) states that are used, along with the change in the configuration coordinate ($\Delta Q$), to evaluate the Huang-Rhys ($S$) factors, which provide an effective average number of phonons created during a vertical transition.\cite{Alkauskas:2014kk,Alkauskas:2016bf,Alkauskas:2016kf} 
The formalism associated with CCD analysis provides insight into whether the defects may contribute to radiative recombination or non-radiative carrier capture with multi-phonon emission, and naturally identifies the energies required to induce charge fluctuations of the localized defect states.\cite{Alkauskas:2014kk,Alkauskas:2016bf,Alkauskas:2016kf} 
The generalized coordinate largely reflects the bond lengths between the DB atom and its nearest neighbors, but also includes a degree of displacement of the DB atom from the plane formed by its three nearest neighbors; the  large atomic displacement of the DB atom associated with different occupancies of the DB state results in fairly large atomic relaxations that lead to large $R^q_{\rm FS}$ relative to the band gap.
As seen within Fig.~\ref{FIG:CCD}a, we find the relaxation energies to be 0.48~eV for the excited acceptor state and 0.45~eV for the excited neutral state configurations of the Ge DB, which yields $\epsilon^o(0/-)$ and $\epsilon^o(-/0)$ values of 0.17~eV and 1.10~eV above the Si VBM, for the acceptor and neutral states, respectively. 
The large relaxation energies and $\Delta Q$ values of the Ge DBs result in large Huang-Rhys factors of $\sim$30 for the excitations in Fig.~\ref{FIG:CCD}.
Considering transitions to both the VBM and CBM in Fig.~\ref{FIG:CCD}a-b, we discuss below our findings of very small classical energy barriers, which facilitate non-radiative capture or emission processes.

\begin{figure}
{\includegraphics[width=1\textwidth]{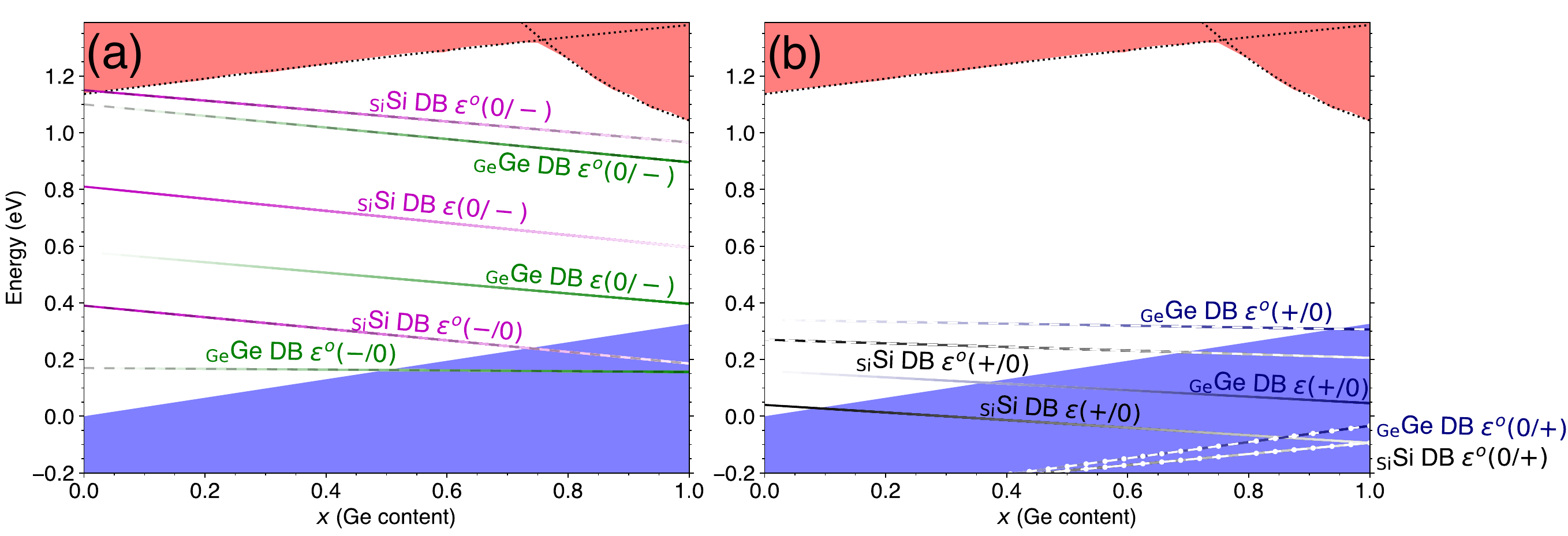}}
\caption{Plot of the calculated thermal (adiabatic, $\epsilon$) and optical (vertical, $\epsilon^o$) transition levels between the neutral, acceptor and donor states of the (a) Si DBs and (b) Ge DBs shown as a function of composition.  
The first (second) charge in the $\epsilon^o$ labelling represents the initial (excited) state.
Shaded regions indicate interpolated values for different back-bonding environments, adopting the same conventions and symbols as in Fig.~\ref{FIG:DB1}. 
We only include vertical transitions that we were able to calculate that exhibit localized defect states; for example all Ge DB-based $\epsilon^o(0/+)$ levels were found to be delocalized in the valence band for all compositions and are thus excluded from Fig.~\ref{FIG:DB2}.
}
\label{FIG:DB2}
\end{figure}

For the Si DB, we find slightly lower relaxation energies of 0.34~eV and 0.42~eV, placing the $\epsilon^o(0/-)$ levels 0.47 and 1.15~eV above the strained Si VBM, for the acceptor and neutral states, respectively (using the 0.85~eV $\epsilon(0/-)$ value from Table~\ref{TAB:EPS}).
As the Huang-Rhys factors scale as $\Delta Q^2$ and linearly with $\Omega$,\cite{Alkauskas:2014kk,Alkauskas:2016kf} the smaller relative lattice relaxations and $\Delta Q$ values associated with Si DB charge excitations compared Ge DBs overcompensate the slightly larger $\Omega$ values and lead to $S$ values approximately half as large as the Ge DBs. 
In general, we find the $R^q_{\rm FS}$ to be within $\sim$0.2-0.4~eV for all Si and Ge DBs across the entire Si$_x$Ge$_{1-x}$ composition range, which leads to excited state $\epsilon^o$ transition levels that shift up and down relative to the thermodynamic transition levels and are summarized in Fig.~\ref{FIG:DB2}, again linearly interpolated across compositions as in Fig.~\ref{FIG:DB1}. 
Coupled with the vibrational energies ranging from 15-25 meV, these generally lead to Huang-Rhys factors in the range of $\sim$15-35 for the DB defect excitations in the Si$_x$Ge$_{1-x}$ alloy.

We find very small classical energy barriers for the excited state of the acceptor to trap a hole and form a neutral state, as seen in Fig.~\ref{FIG:CCD} for $_{\rm Si}$Ge$_{\rm Si}$ DBs in unstrained (a) and strained Si (c).
The calculated barrier for hole capture is 0.02~eV for the $_{\rm Si}$Ge$_{\rm Si}$ DB (Fig.~\ref{FIG:CCD}a) and 0.07~eV for the Si DB (not shown), with the values marginally lowered in strained Si (0.01~eV and 0.07~eV, for Ge and Si DBs, respectively).
Additionally, a population of neutral DBs can capture electrons from the CBM to form an excited state acceptor, which we find exhibit even lower classical barriers as illustrated in Figs.~\ref{FIG:CCD}(b) and (d). 
From Fig.~\ref{FIG:CCD}(d), it can be seen that the effect of strain on reducing the band gap favors non-radiative over radiative transitions owing to the intersection of potential energy surfaces at intermediate configuration coordinates.  
This also holds true for Si DBs in strained Si, which has a slightly larger calculated classical barrier (0.03~eV) relative to the 0.01~eV for Ge. 
Considering that neutral DBs are expected to be the dominant population in nominally undoped Si, any Si or Ge DBs may readily trap free carriers that are either thermally generated or produced during operation, with the effects expected to be most pronounced in the strained layers.
These results identify Ge DBs may be more prone to trapping events that could cause fluctuating charges, particularly in the strained Si/SiGe heterostructures, and indicate that improperly passivated heterostructure interfaces will more readily exhibit these types of excitations.
While a formal quantitative treatment of the carrier capture rates requires higher level of theory,\cite{Alkauskas:2014kk} these classical barriers suggest a facile exchange of charge in DBs that may indicate these types of defects will contribute to charge noise in Si/\sige quantum dots and other related devices. Such small barriers coupled with the observed influence of the local environment of the DB energy landscape through band edge, compositional, and strain fluctuations likely leads to a distribution of barriers that may all contribute to charge-state excitations at different rates.

\subsubsection{Dangling bonds in oxidized layers}
Oxide gate dielectric layers were recently linked with charge noise in Si/\sige quantum dot  devices,\cite{Connors:2019ej} which suggests that dangling bonds in the vicinity of the oxide interface(s) may be among the most significant sources of charge noise.
We consider the calculated levels of the Si and Ge dangling bonds in SiO$_2$ as a proxy for the positions of these levels in oxidized layers that either can form on strained Si capping layers in many Si/\sige quantum dot architectures or are present in MOS-type architectures.\cite{Simmons.2009.10.1021/nl9014974,Simmons.2007.10.1063/1.2816331,Rudolph.2019.10.1038/s41598-019-43995-w} 
The general family of DB defects in SiO$_2$ has been extensively studied in the past (e.g. $E^\prime$ centers in amorphous silica) owing to the extreme technological importance of the native oxide/semiconductor interface in Si.\cite{Skuja:1998fx, Pacchioni:1998bw,ElSayed:2013hr, Ling:2013ko, ElSayed:2015gy, Adelstein:ui}
We note that $E^\prime$ centers are associated with O deficiency in SiO$_2$ that results in unpaired Si $sp^3$ DBs that can exhibit distinct charge and spin configurations. 
There is less work on the Ge analogs to $E^\prime$ centers, e.g. Ge$_{\rm Si}$ back-bonded to 3 O atoms in the SiO$_2$ matrix, but some evidence exists that the levels are located at similar energetic positions as the Si DBs.\cite{Skuja:1998fx}

\begin{figure}
\includegraphics[width=1.0\textwidth]{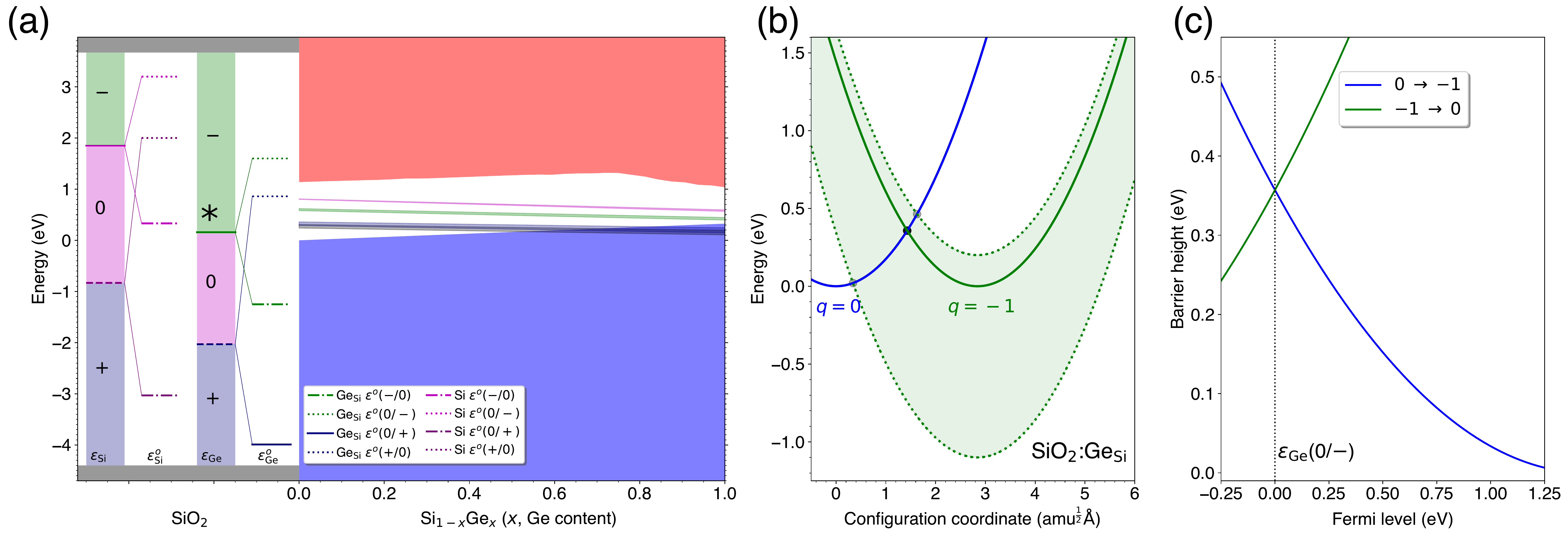}
\caption{ (a) Plot of the calculated thermal (adiabatic, $\epsilon$) and optical (vertical, $\epsilon^o$) transition levels for Si and Ge$_{\rm Si}$ DBs shown in SiO$_2$, and also including the band alignment with \sige{} as shown in Fig.~\ref{FIG:DB1}. The first (second) charge in the $\epsilon^o$ labelling in the legend represents the initial (excited) state. (b) Calculated CCD for electron capture (emission) the neutral (negatively-charged) Ge$_{\rm Si}$ DB in SiO$_2$ associated with the $\epsilon$(0/$-)$ excitation of the Ge$_{\rm Si}$ DB in (a) which is highlighted with an asterisk. The energy of the Ge$_{\rm Si}^{-}$ DB is shown over a range of Fermi levels corresponding to the Si VBM to CBM relative to the $\epsilon_{\rm Ge}$(0/$-)$, with the classical barrier height highlighted with black and grey points. (c) Variation of the classical barrier height with the Fermi level relative to the $\epsilon_{\rm Ge}$(0/$-)$ in SiO$_2$, spanning from the Si VBM to CBM that spans the full range of \sige. } \label{FIG:DBOXIDE}
\end{figure}


We find the Si DB has a deep $\epsilon(+/0)$ transition 3.57~eV above the VBM, with the $\epsilon(0/-)$ falling 6.25~eV above the VBM, as summarized Table~\ref{TAB:EPS}. Considering an experimental VBO of $\sim$4.4~eV for the Si/SiO$_2$ interface,\cite{ElSayed:2015gy,Alkauskas:2008ej} this suggests the Si DB in SiO$_2$ will be neutral for all Fermi levels spanning $p$ to $n$-type Si, with all transition levels far from the Si band edges. Interestingly, we find that the Ge$_{\rm Si}$ DB in SiO$_2$ is not as deep and exhibits the $\epsilon(0/-)$ transition 4.56~eV above the SiO$_2$ VBM, which is aligned $\sim$0.1-0.2~eV above the VBM of Si, based on the experimental band offsets. These results are depicted in Fig.~\ref{FIG:DBOXIDE}, which includes the calculated defect levels of the Si and Ge$_{\rm Si}$ DBs in SiO$_2$ relative to the band edges of Si$_x$Ge$_{1-x}$. 
If we also consider the vertical excitation energies, we find that some of these excited states can fall within the band gap of Si owing to the large Franck-Condon shifts ranging from $\sim$1.4-2.8~eV in the highly ionic SiO$_2$, as also shown in Fig.~\ref{FIG:DBOXIDE}. 
While most excitation processes associated with DBs in SiO$_2$ are expected to yield optical signatures from radiative recombination, we highlight the potential for non-radiative carrier capture or emission at neutral or negatively-charged Ge$_{\rm Si}$ DBs in Fig.~\ref{FIG:DBOXIDE}. 
We show in Fig.~\ref{FIG:DBOXIDE}b and c, that depending on the Fermi level in the \sige material, the classical barriers for electron capture and emission are 0.36 eV for a Fermi level position corresponding to the $\epsilon(0/-)$ for the Ge DB in SiO$_2$.
This scenario could be relevant for more $p$-type Si layers, where any Ge$_{\rm Si}$ DBs present within surface oxides would preferentially be in the neutral charge state and could trap electrons, also leading to charge fluctuations between the neutral and negatively-charged acceptor states.  
For higher Fermi levels corresponding to more $n$-type Si$_x$Ge$_{1-x}$, the acceptor state of the DB would become more favorable and the relevant barriers for electron emission increase (green lines in Fig.~\ref{FIG:DBOXIDE}c).
Overall, these results indicate Ge incorporation in oxidized Si surface layers may be more problematic from the standpoint of fluctuating charge traps, as these levels may be more readily ionized for Fermi levels within the Si band gap and closer to the VBM. Additionally, the large relaxation energies and ionic displacements associated with these capture events lead to an extremely large Huang-Rhys factor of $\sim$40, indicating a large number of phonons are expected to be emitted in this process.

\section{Conclusions}

In conclusion, we evaluated a series of dangling-bond-type defects associated with Si and Ge in Si, Ge, Si$_x$Ge$_{1-x}$, and SiO$_2$ to assess their possible roles in contributing charge noise in Si/SiGe quantum dot qubits, via charge fluctuations associated with electron and hole capture and emission processes. 
We employed hybrid density functional theory calculations to study the thermodynamic and optical charge-state transition levels of these defects in Si$_x$Ge$_{1-x}$ and SiO$_2$, and find an overall good agreement of calculated values with available experimental data for both qualitative and quantitative assessment of these defects. 
We find that DBs exhibit donor and acceptor levels that span a considerable energy range in Si and Si$_x$Ge$_{1-x}$ alloys that depends on the composition and local environment of the defects. 
Overall, our results indicate that Ge DBs tend to exhibit a greater tendency to form over Si DBs and exhibit lower classical barriers for non-radiative recombination compared to Si DBs if they are present in Si$_x$Ge$_{1-x}$ or in oxidized layers. The span of energies associated with Ge DBs are also found to be more sensitive to their local environment than Si.
This suggests that compositional variations, strain, and modulation of the Fermi level can lead a range of energies for these states, not only affecting the appearance of charge trapping levels, but also causing a large variation in excitation energies and transition/fluctuation rates that could contribute to a 1/$f$ frequency dependence commonly observed in qubit devices.\cite{Rudolph.2019.10.1038/s41598-019-43995-w}


We find small classical barriers for these defects on the order of 10s of meV, suggesting facile charge exchange processes. When coupled with the observed influence of the local environment of the DB energy landscape through fluctuations in the band edge positions with composition and strain, likely leads to a distribution of barriers that may all contribute to charge-state excitations/fluctuations at different rates that overlap with the operational frequencies in these quantum dot devices. In addition, we find that oxygen deficiency-type defects in Si or \sige oxides can also trap and exchange carriers from the adjacent semiconductor, leading to charge fluctuations between neutral and positively-charged states. Germanium incorporation in oxidized Si layers was found potentially to be most problematic from the standpoint of fluctuating charge traps, as the associated Ge DB levels in the oxide may be more readily ionized under device conditions.
These results suggest that DB-type defects may contribute to charge noise across a wide frequency spectrum in Si/\sige quantum dot devices and emphasizes the importance of optimizing growth processes of the graded \sige structures and possible advantages of employing passivation schemes to mitigate these types of defects.


\begin{acknowledgement}
We thank Mark Friesen and Mark. A. Eriksson for useful discussions.
This work was performed under the auspices of the U.S. Department of Energy at Lawrence Livermore National Laboratory under Contract DE-AC52-07NA27344.
Work by V.L. was supported by the U.S. Department of Energy, Office of Science, Basic Energy Sciences, Materials Sciences and Engineering Division.
LLNL-JRNL-849060
\end{acknowledgement}

\bibliography{DRAFT-qubits_ACS_v5b.bbl}

\end{document}